\documentclass[preprint2]{aastex6}

\shorttitle{Gender Systematics in Telescope Time Allocation at ESO}
\shortauthors{Ferdinando Patat}

\begin{document}


\title{Gender Systematics in Telescope Time Allocation at ESO}


\author{Ferdinando Patat\altaffilmark{1}}
\affil{European Southern Observatory\\
K. Schwarzschildstr. 2 \\
D-85748 Garching b. M\"unchen, Germany}


\altaffiltext{1}{fpatat@eso.org}

\begin{abstract}

The results of a comprehensive statistical analysis of gender systematics in the time allocation process at European Southern Observatory (ESO)  are presented. The sample on which the study is based includes more than 13000 Normal and Short proposals, submitted by about 3000 principal investigators (PI) over eight years. The genders of PIs, and of the panel members of the Observing Programmes Committee (OPC), were used, together with their career level, to analyse the grade distributions and the proposal success rates. Proposals submitted by female PIs show a significantly lower probability of being allocated time. The proposal success rates (defined as number of top ranked runs over requested runs) are 16.0$\pm$0.6\% and 22.0$\pm$0.4\% for females and males, respectively. To a significant extent the disparity is related to different input distributions in terms of career level. The seniority of male PIs is significantly higher than that of female PIs, with only 34\% of the female PIs being professionally employed astronomers (compared to 53\% for male PIs). A small, but statistically significant, gender-dependent behaviour is measured for the OPC referees: both genders show the same systematics, but they are larger for males than females. The PI female/male fraction is very close to 30/70; although far from parity, the fraction is higher than that observed, for instance, among IAU membership. 

\end{abstract}

\keywords{sociology of astronomy -- history and philosophy of astronomy}

\submitted{Draft version \today}



\section{Introduction} \label{sec:intro}

The ESO 2020 prioritisation exercise (\cite{primas}) spawned a number of actions. Among them was the constitution of a Time Allocation Working Group (TAWG), which has been tasked with the review of the whole telescope time allocation process at the European Southern Observatory (ESO). The TAWG, chaired by the author of this article, will submit a set of recommendations to the Director for Science, to be presented to the ESO Scientific Technical Committee in October 2016. The activities of the TAWG included a wide range of statistical studies on time requests, proposal grading and possible trends in the allocation process. The results will be presented in a separate report, while this article focuses solely on the gender aspect.

Obtaining telescope time at world-leading facilities is fundamental for the well-being of an astronomer's scientific activity. Therefore, systematics in the time allocation process can have negative consequences on the career of a researcher. The present study was carried out along the lines traced by \cite{reid}, who conducted a similar analysis for the Hubble Space Telescope (HST) proposal selection process. The main purpose is the quantification of gender dependent systematics and subsequent considerations in the wider context of possible correlations between the final merit attributed to a proposal and aspects that are scientifically irrelevant.

\begin{figure*}
\begin{center}
\includegraphics[width=14cm]{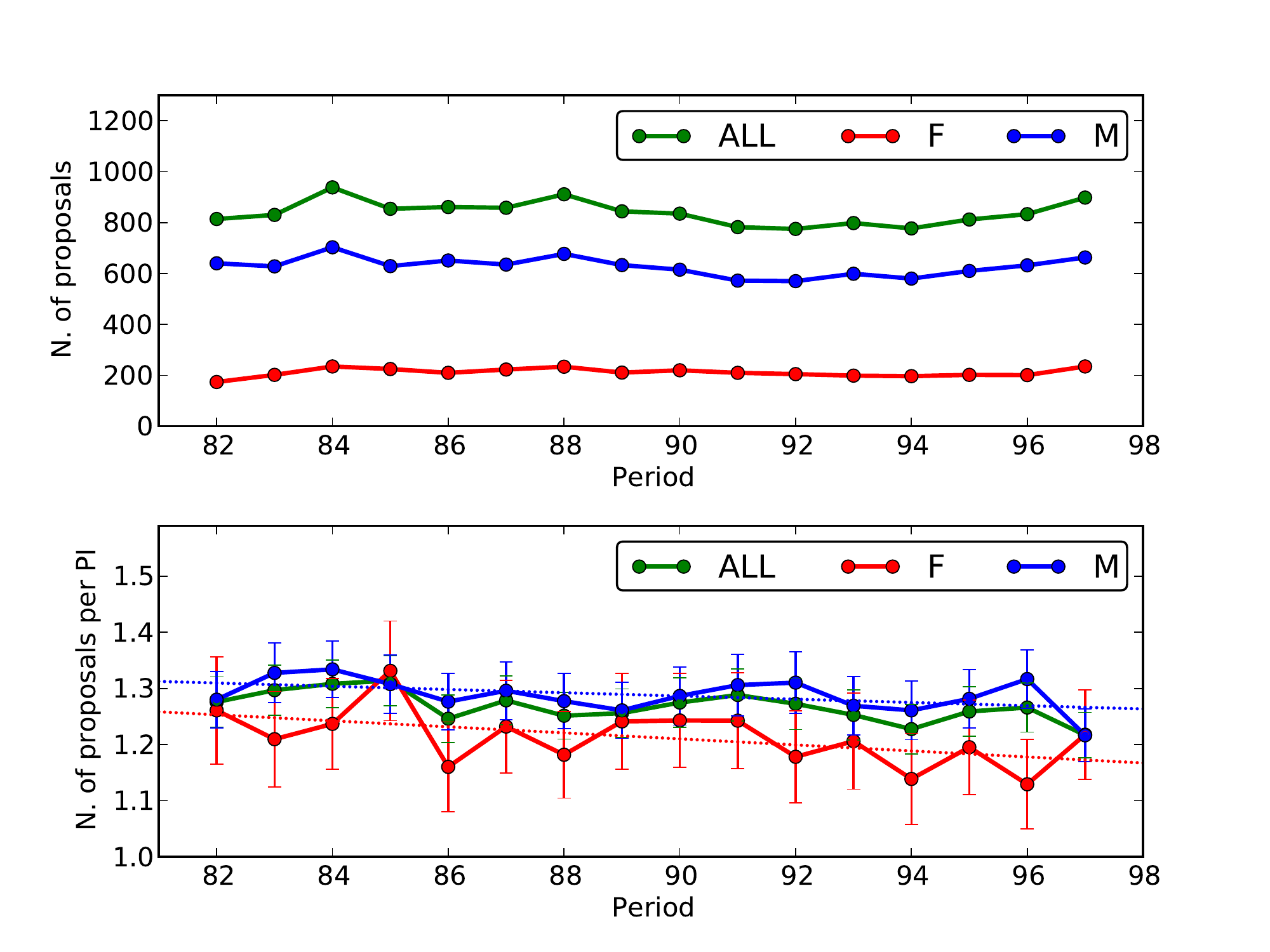}
\caption{\label{fig:fig1}Trends of proposal submission over Periods 82-97. Upper panel: number of submitted proposals. Lower panel: number of proposals per PI. In this and all following plots, the error-bars indicate the Poisson noise, while the dotted lines trace a linear least squares fit to the data points.}
\end{center}
\end{figure*}

\section{\label{sec:review}The ESO proposal review process}

The proposal review process currently deployed at ESO is described in detail in \cite{patat}. Here only a brief summary will be provided. The review is performed in two steps, before and during the Observing Programmes Committee (OPC) meeting, indicated as pre- and post-OPC.
During the pre-OPC phase the proposals assigned to a given panel are read and graded by all non-conflicted panel members. Each panel is composed of six referees. The grade scale is between 1 (best) and 5 (worst), where a grade larger than 3 will not be considered for scheduling. The pre-OPC grade is entered into the database independently and secretly by each panel member. Every proposal is assigned a primary referee, who will present the case during the panel discussion and be in charge of editing the comment sent to the principal investigator (PI). No special weight is attributed
to the primary referee's grade. 

A proposal may contain one or more runs, which
are graded separately by the reviewers. The final grade of a run is computed as the average of the grades given by all non-conflicted panel members. In order to account for systematic differences between the various referees, before computing the final average the grade distributions of the reviewers are normalised to have the same average and standard deviation.

Once the pre-OPC grading is complete, the Observing Programmes Office (OPO) compiles ranked lists per telescope and applies triage, removing the bottom 30\% (by observing time). With very few exceptions (at the discretion of the panels), triaged proposals are not discussed during the OPC meeting and their fate is fully dictated by the pre-OPC grades. The surviving proposals are discussed by all non-conflicted panel members during the OPC meeting and graded (again secretly). Unlike in the pre-OPC phase, the post-OPC grades are logged anonymously in the database (meaning that
the referee's identity is lost). Once all panels have completed their activities, the ranked lists per telescope are computed by OPO using the average run grade, after normalising the grade distributions of the panels to have the same average and standard deviation.
The ranked lists are finally used to schedule the various telescopes. 

On account of the different oversubscription rates and different demands in terms of right ascension and observing conditions,
two runs having the same OPC grade may end up having different allocation outcomes.
For any given telescope, the rank class A is assigned to the top 50\% (by time)
of the Service Mode (SM) runs that were allocated time, while the remaining runs are assigned to rank B. The grade at which the A/B transition occurs changes from telescope to telescope, depending on the demands and the exact time distribution of allocated runs. No priority rank is assigned to Visitor Mode (VM) runs, whose observations, once scheduled, are conducted in any case, barring adverse weather conditions. As the grading takes place by run, the statistics presented in this study refer to runs and not to proposals.

\begin{table*}
\caption{\label{tab:tab1}Distribution of referees (upper), PIs (middle) and proposals (lower) by career level and gender.}
\tabcolsep 4mm
\begin{tabular}{lccccccccccc}
\multicolumn{11}{c}{{\bf Referees}}\\
\hline
Career level & \multicolumn{2}{c}{All} & \multicolumn{3}{c}{F} & \multicolumn{3}{c}{M} & \multicolumn{2}{c}{Balance}\\
 & N & \% & N & \% of all & \% of F & N & \% of all & \% of M & F & M \\
\hline
Astronomer & 462 & 87.7 & 129 & 24.5 & 83.2 & 333 & 63.2 & 89.5 & 27.9 & 72.1 \\
Post-doc     & 65  & 12.3 & 26 & 4.9 & 16.8 & 39 & 7.4 & 10.5 & 40.0 & 60.0 \\
\hline
All               & 527  &  & 155 & 29.4 & & 372 & 70.6 & & & \\
 & & & & & & & & & & \\
 \multicolumn{11}{c}{{\bf Principal Investigators}}\\
 \hline
 Career level & \multicolumn{2}{c}{All} & \multicolumn{3}{c}{F} & \multicolumn{3}{c}{M} & \multicolumn{2}{c}{Balance}\\
& N & \% & N & \% of all & \% of F & N & \% of all & \% of M & F & M \\
\hline
 Astronomer & 1418 & 47.4 & 294 & 9.8 & 33.9 & 1124 & 37.6 & 52.9 & 20.7 & 79.3 \\
 Post-doc     & 1013 & 33.9 & 366 & 12.2 & 42.3 & 647 & 21.6 & 30.5 & 36.1 & 63.9 \\
 Student       & 559 & 18.7 & 206 & 6.9 & 23.8 & 353 & 11.8 & 16.6 & 36.9 & 63.1 \\
 \hline
 All                & 2990 & & 866 & 29.0 & & 2124 & 71.0 & & & \\
 & & & & & & & & & & \\
 \multicolumn{11}{c}{{\bf Proposals}}\\
 \hline
 Career level & \multicolumn{2}{c}{All} & \multicolumn{3}{c}{F} & \multicolumn{3}{c}{M} & \multicolumn{2}{c}{Balance}\\
& N & \% & N & \% of all & \% of F & N & \% of all & \% of M & F & M \\
\hline
Astronomer & 7516 & 56.4 & 1377 & 10.3 & 41.2 & 6139 & 46.1 & 61.5 & 18.3 & 81.7 \\
Post-doc     & 4595 & 34.5 & 1522 & 11.4 & 45.5 & 3073 & 23.1 & 30.8 & 33.1 & 66.9 \\
Student & 1219 & 9.1 & 446 & 3.3 & 13.3 & 773 & 5.8 & 7.7 & 36.6 & 63.4 \\
\hline
All & 13330 & & 3345 & 25.1 & & 9985 & 74.9 & & & \\
\end{tabular}
\end{table*}

\section{\label{sec:data}Characteristics of the proposal dataset}

The data on which this study is based were extracted from the ESO proposal database. The sample includes only Normal and Short proposals. Large Programmes, Calibration Programmes and Surveys were not included because they are not graded by the panels. Guaranteed Time Observations (GTO), Target of Opportunity (TOO) and Monitoring proposals were also excluded because of their special nature, which would introduce systematic effects in the statistics. Normal and Short programmes account for more than 85\% of the runs requested every semester.

The database contains pre- and post-OPC proposal gradings starting from Period P79 (April 2007-September 2007). However, the data are properly and consistently stored only from P82 (October 2008-March 2009). For this reason, the analysis presented here covers ESO Periods 82 to 97 (April 2016-September 2016), inclusive. This eight-year interval can be considered representative of regular operations, with the Very Large Telescope (VLT) and Very Large Telescope Interferometer (VLTI) in full activity and with practically every telescope focus occupied by an instrument. In addition, during the above period the ESO User Portal (UP) was fully functional, so that PI information could be retrieved in a consistent, homogeneous and complete way.
In Periods 82 and 83, the OPC included 12 panels. An extra panel was added to category A (Cosmology) in P84. From P84 onwards, the OPC had a stable composition of 13 panels, with six members each. The panel members serve for typically two semesters, although in order to maintain some memory, a fair number of them are asked to stay for an extra semester. The OPC-proper members (i.e., panel chairs and members-at-large) normally serve for four semesters. The full dataset includes 22022 runs (13420 proposals) submitted by 3017 distinct PIs (about 4.4 proposals per PI, at an average rate of $\sim$840 proposals per semester). About 65 \% of the proposals include one single run, while $\sim$85\% of the proposals have fewer than three runs. The runs were reviewed by 527 distinct referees, who assigned 123 358 pre-OPC grades and 100558 post-OPC grades. On average, each proposal was reviewed by 5.6 referees, with about 95\% of the proposals reviewed by 5 or 6 referees.

\begin{figure}
\includegraphics[width=9cm]{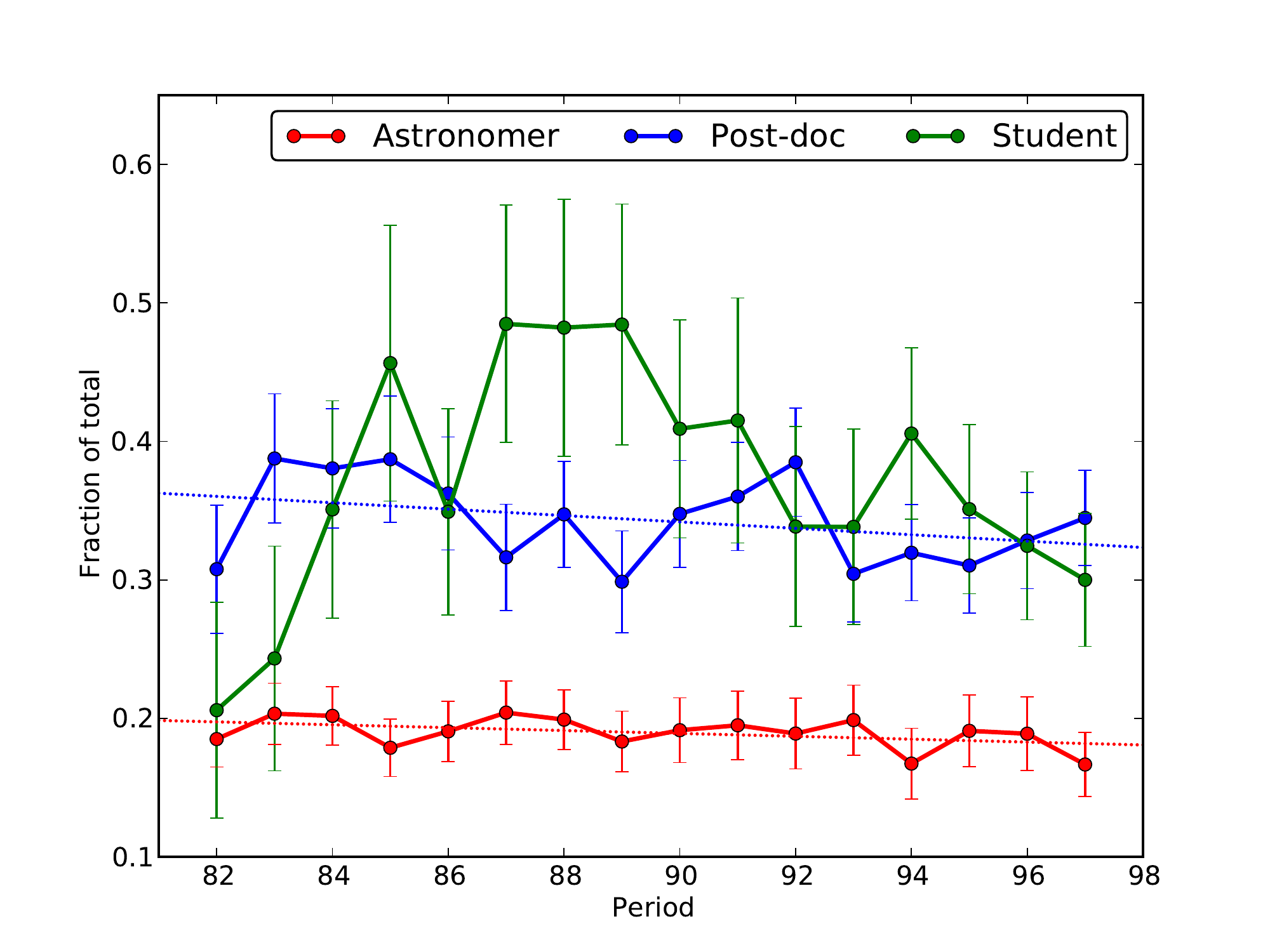}
\caption{\label{fig:fig2}Evolution of the fraction of female PIs by career level.}
\end{figure}
\section{\label{sec:career}Career level and gender of proposers and referees}

All UP subscribers can set their career level, choosing from one of the following options: Professional astronomer (hereafter AP), Teacher/Educator/Journalist, Post-doc in astronomy (PD), Student (ST), Amateur astronomer, Non-astronomer scientist, or Other. It is important to note that, like other information stored in the UP, this does not necessarily reflect the actual status at the time of proposal review or submission, and represents a snapshot relative to the last UP profile update. Therefore, there can be cases in which a user created the account when she/he was a student and then never updated the profile, although she/he may now be a post-doc or even a professional astronomer. In the extreme scenario in which no users update their career level after the creation of their UP account (for PIs most likely when they submitted their first proposal), one expects that the fractions of ST and PD scientists are overestimated, while that of AP is under-estimated.
This effect would produce time trends, with the fraction of AP PIs steadily decreasing with time, which is actually what is seen in the data. This needs to be kept in mind when considering the accuracy of the results related to the career level. For our purposes, only AP, PD and ST levels are relevant, as they account for 99.0\% of PIs; the remaining categories were ignored. The overall fractions of PIs in the three classes are: 47.4\% (AP), 33.9\% (PD) and 18.7\% (ST). By construction, the OPC panel composition is heavily dominated by professional astronomers (87.7\%), while the small remaining portion is constituted by post-docs (12.3\%).

The referee and PI genders are not stored in the UP. For this reason, the gender had to be deduced through personal knowledge, from the first name in email addresses and through web search, similar to what was done by \cite{reid}. For the 527 distinct referees, there is confidence that the gender classification is exact for all entries. For the $\sim$3000 distinct PI classifications, the confidence level is estimated to be better than 99\%. The gender could not be determined for five PIs, who were excluded from the analysis. The final set of PIs belonging to one of the three career levels AP, PD or ST, for which gender information is available, includes 2990 scientists, who submitted 13330 proposals.
The distributions per career level and gender are summarised in Table \ref{tab:tab1} for referees (top), PIs (middle) and submitted proposals (bottom); for proposals, career level and gender refer only to the PI. 

The first aspect worth noting is that the overall female (F) and male (M) fractions are close to 30/70 in both the referees and the PIs division. The observed differences are within the expected Poisson uncertainties, so that the F/M proportion in the panels can be considered identical to that of the PI community. Of course, the gender fraction in the PI sample does
not necessarily reflect that of the overall astronomical community, as there may be selection effects relating to gender when it comes to choosing the leader of a proposal. The magnitude of these effects could be estimated by computing the ratio by gender between the number of PIs and total applicants. This analysis, which may reveal an additional selection bias at source, is not possible at the moment as the gender of the proposal co-investigators is not available.

The second aspect, which is very important for the purposes of this analysis, is the diversity of the career level fraction by gender. In general, both for the referees and the PIs, the average professional seniority level is higher for male scientists. This difference is particularly marked for PIs, for which the number of professional astronomers is 19\% larger for M than for F applicants (see Table \ref{tab:tab1}). Since, at least to some extent, the quality of a proposal is expected to grow with the PI's career level\footnote{This correlation is less obvious than may be expected. For instance, the fact that the PI of a proposal is a student does not necessarily mean the proposal was written entirely by the student. It is plausible that the supervisor had a role, and contributed in a significant way, both in terms of content and presentation. On the other hand, it is also reasonable to expect that some supervisors, for educational purposes, leave some degree of independence to the student. In addition, the choice of the PI may be also dictated by strategic arguments within the proposing team, not necessarily and strictly related to science and/or who wrote the proposal. All these aspects contribute to blurring the possible correlation between PI career level and scientific merit of the proposal.}, it is to be expected that the depletion of professional astronomers as female PIs turns into a lower success rate. This will be discussed in more detail below. As a by-product of this analysis, one can also look at the evolution of the PI gender balance as a function of professional career level (see Table \ref{tab:tab1}, last two columns). While there is no statistically significant evolution between the ST and PD levels (with a F/M balance $\sim$36/64), a clear jump is seen in the AP level ($\sim$21/79). If the PI community appears already to be unbalanced from the career start, the situation clearly degrades in the last stage. The lack of a finer level resolution (both within the PD and the AP classes) does not allow any more refinement in the study of the career trajectory by gender.
The diversity seen in the PIs is reflected in the distribution of submitted proposals by gender and career level (Table \ref{tab:tab1}, lower). The numbers are similar, but all differences tend to become more pronounced when considering the proposals. For instance, the overall gender fraction is 25/75 (vs. 29/71), and the AP gender fraction is 18/82 (vs. 21/79). In addition, the male career level distribution becomes more top-heavy (61.5\% of the proposals are submitted by senior astronomers) than for female PIs (41.2\%). The fraction of proposals submitted by female professional astronomers (41.2\%) is comparable to that of post-docs (45.5\%), while for male PIs, the AP fraction (61.5\%) is about a factor of two more than the PD fraction (30.8\%).

\begin{figure}
\includegraphics[width=9cm]{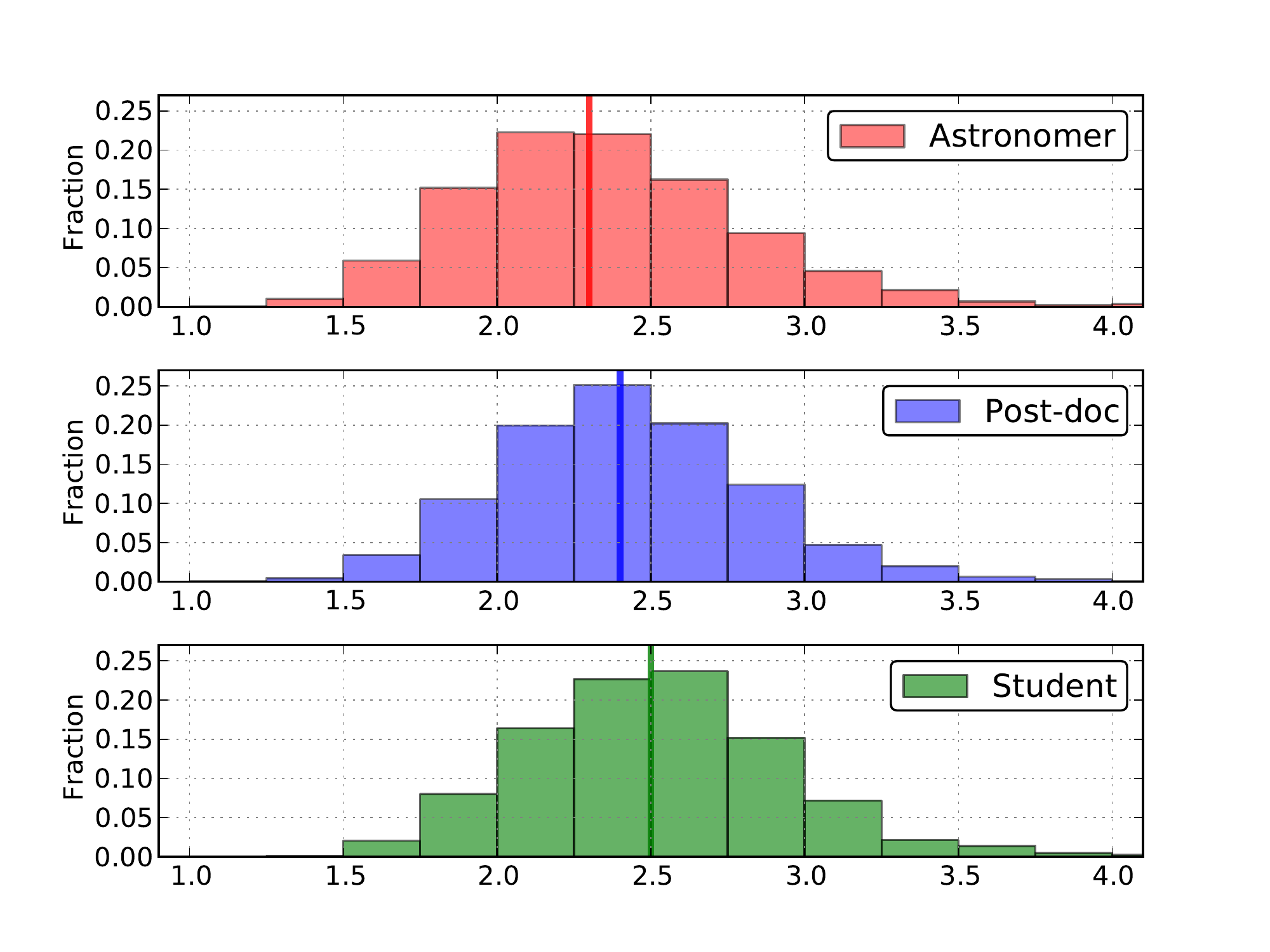}
\caption{\label{fig:fig3}Pre-OPC average grade distributions by PI career level: Professional astronomer (upper), Post-doc (middle) and Student (lower). The vertical lines indicate the average values of the three distributions.}
\end{figure}

\section{Trends in proposal submission}

The proposal submission trend by PI gender is presented in Figure \ref{fig:fig1} (upper panel). The overall average submission rate (proposals per semester) is 838.8$\pm$11.5, while it is 211.4$\pm$4.0 (25.2\%) and 627.3$\pm$8.9 (74.8\%) for F and M PIs, respectively. The number of proposals
is particularly stable for F PIs. In addition to the difference in the F and M fractions (which are directly related to the PI gender distribution, see below), it is worth noticing that the two genders differ significantly in their submission rate in terms of proposals per semester per distinct
PI (Figure \ref{fig:fig1}, lower panel). The overall average rate is 1.27$\pm$0.01 proposals per PI, and shows a mild decrease during the period covered by this study. The submission rate for M PIs (1.29$\pm$0.01) is larger than that of F PIs (1.21$\pm$0.01): on average, a female PI submits $\sim$8\% fewer proposals per semester. There is some evidence for a steady decrease of the submission rate with time. 

The number of proposals remained roughly constant during the period under consideration (see Figure \ref{fig:fig1}). This steady situation is also common to the number of distinct PIs per semester. The overall average number of PIs is 661.8$\pm$9.5 (standard error of the mean), while for F and M the average is 174.5$\pm$3.3 and 487.0$\pm$7.5, respectively. Every semester only a fraction of all potential PIs submit
a proposal\footnote{The total number of distinct investigators (PIs and co-Is) per semester steadily evolved from ~$\sim$2500 to $\sim$3500 across the considered time span. The total number of distinct investigators exceeds 10000; if all these are considered as potential PIs, about two thirds never submitted a proposal over the eight years covered by this study.}. The data show that, on average, every semester the list of distinct
PIs includes about half of those who submitted a proposal in the previous period. This fraction is larger for M PIs ($\sim$55\%) than for F PIs ($\sim$40\%), implying that the ratio between the number of effective PIs and the number of potential PIs is smaller for F than for M scientists (for whom the opposite trend is observed).
The fraction of F PIs increased by a few percent over the eight years considered in this report, with an average value around 26\%. A closer look reveals that the period-by-period fraction is significantly smaller than the global sample value deduced considering all periods together (28.8\%). This is explained by the effect described above, which is related to the different behaviour of the female astronomer community in terms of period-to-period changes in the PI set. The gender fractions within the three career levels are roughly constant in time, as illustrated in Figure \ref{fig:fig2} for the F PIs (the larger fluctuations seen for AP, and especially ST, PIs are due to the small number statistics. The best fit line for the ST class was omitted because the slope is characterised by a large uncertainty and is consistent with no time evolution).

\begin{figure}
\includegraphics[width=9cm]{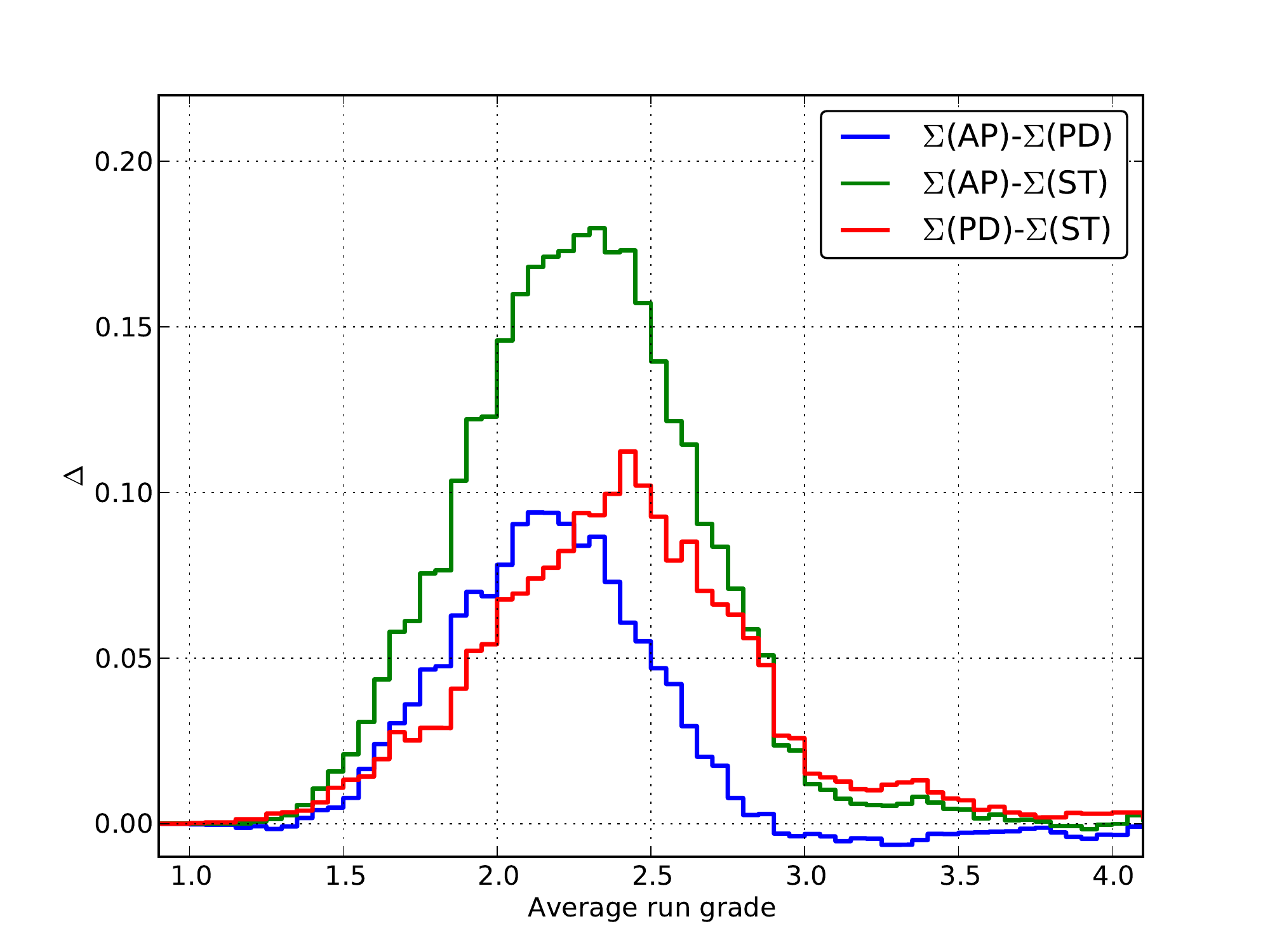}
\caption{\label{fig:fig4}Differences between the cumulative functions of the pre- OPC run grade distributions.}
\end{figure}

\section{Career level, gender and proposal grading}

\begin{figure*}
\begin{center}
\includegraphics[width=14cm]{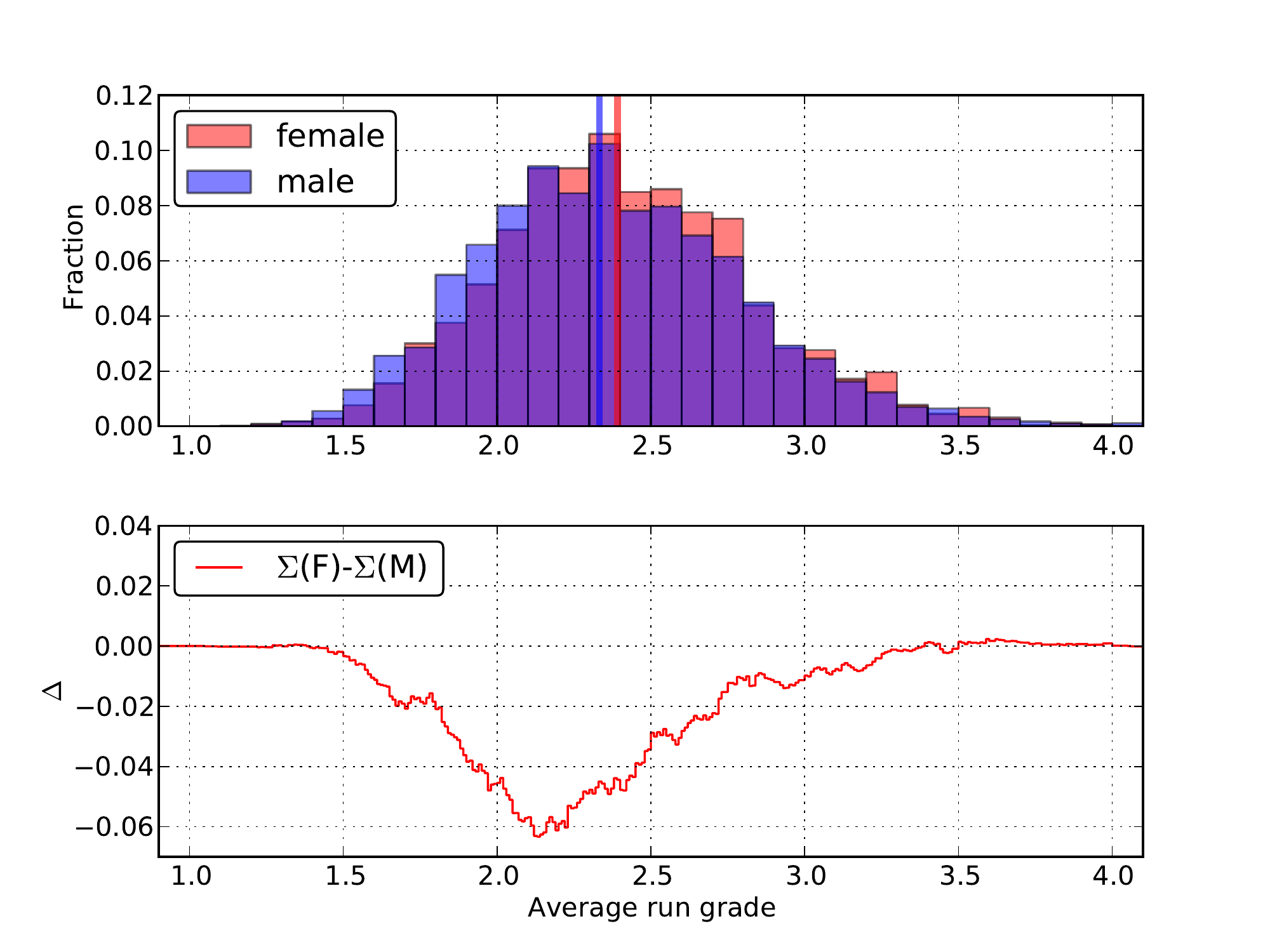}
\caption{\label{fig:fig5}Upper panel: Professional astronomers and Post-doc pre-OPC run grade distributions for M (blue) and F (red) PIs. Lower panel: difference between the cumulative distribution functions (F-M).}
\end{center}
\end{figure*}

The pre-OPC average grade distributions are presented in Figure \ref{fig:fig3} for the three career levels. A first glance at the plots reveal that there are statistically significant differences, with a progressive shift of the central values towards poorer grades as one proceeds through the AP, PD and ST levels. Classical statistical tests (such as Kolmogorov-Smirnov) show that the three distributions differ at very high confidence levels ($>$99.999\%). The deviations between the distributions can be robustly quantified using the differences ($\Delta$) of their cumulative functions ($\Sigma$), which are plotted in Figure \ref{fig:fig4}. The largest deviation is seen between the AP and ST distributions: the number of runs with grades better than 2.2 is 18.1\% larger for AP than for ST proposals. This difference reduces to 9.9\% when considering AP-PD distributions, and to 11.0\% when considering PD-ST distributions. In the latter case the maximum difference is attained at grade 2.4, which signals a weaker advantage of PD proposals over ST proposals (see Figure \ref{fig4}, red curve).

In light of these results, there is no doubt that reviewers systematically attribute different merit to proposals submitted by PIs at different career levels, with professional astronomers scoring better than post-docs, who in turn score better than students. Of course one can still ask whether the observed differences are intrinsic (more experienced sciefntists write better proposals) or perceived ("this proposal was written by a senior scientist, therefore it must be good"). Irrespective of this consideration, the fact remains that there is a significant difference in the grade distributions, which translates into different probabilities of obtaining telescope time. This is quantified in Table \ref{tab:tab2}, which presents the success rates (scheduled runs over requested runs) by career level, rank class and gender.

\begin{table*}
\caption{\label{tab:tab2}Run success rate by PI career level, rank class and gender (in percent). The values in parentheses indicate the Poisson uncertainty.}
\tabcolsep 5.5mm
\begin{tabular}{lcccccc}
\multicolumn{7}{c}{\bf Gender}\\
  & \multicolumn{2}{c}{All} & \multicolumn{2}{c}{F} & \multicolumn{2}{c}{M} \\
Career level  & A+VM & A+B+VM & A+VM & A+B+VM & A+VM & A+B+VM \\
\hline
Astronomer & 23.4 (0.4) & 36.2 (0.5) & 18.3 (0.9) & 32.5 (1.2) & 24.4 (0.5) & 36.9 (0.6) \\
Post-doc & 18.3 (0.5) & 30.5 (0.6) & 14.5 (0.8) & 29.1 (1.1) & 20.0 (0.6) & 31.1 (0.8) \\
Student & 13.2 (0.8) & 25.0 (1.1) & 12.9 (1.3) & 26.4 (1.9) & 13.5 (1.1) & 24.2 (1.4) \\
\hline
All & 20.7 (0.3) & 33.3 (0.4) & 16.0 (0.6) & 30.5 (0.8) & 22.2 (0.4) & 34.2 (0.5) \\
\end{tabular}
\end{table*}

The success rate of runs of rank A and  VM for professional astronomers (23.4\%) 
is $\sim$1.3 times larger than for post-docs (18.3\%), and $\sim$1.8 times larger than
for students (13.2\%). The statistical significance of the differences exceeds the 5$\sigma$ level. Since the career level distributions of F and M PIs are different, this is expected to have an impact on the
corresponding overall success rates by gender, even in the complete absence
of gender effects in the review process. In this idealised case, the overall success rates by gender can be simply predicted by computing a weighted average of the success rates by career level deduced for the whole sample. The weights are he fractions of runs requested by the two genders in the three career bins (Table \ref{tab:tab1}). This simple calculation leads to A+VM success rates of 19.3\% (F) and 22.1\% (M), which provide an estimate of the success rate produced by a pure difference between the career level mixtures of the two gender sets. These values
can be compared to the measurements presented in the last row of Table \ref{tab:tab2} for
F/M of 16.0\% and 22.2\%, respectively. While the M values are fully compatible,
the F rate is smaller, signalling an additional effect. This may be explained as a
combination of too coarse a career level granularity (for instance, the AP class
ranges from an entry-level lecturer to a full professor) and/or genuine gender systematics in the review process.

An indication of the presence of the granularity problem comes from the fact
that the advantage of M PIs does not significantly decrease when looking at
single career levels. With the exception of the ST class, for which the M and
F values are indistinguishable within the noise (see Table \ref{tab:tab2}), the A+VM success
rate ratios are 1.33$\pm$0.07 and 1.38$\pm$0.09 for the AP and PD levels, respectively. This leads to the legitimate suspicion that a finer classification would reveal further gradients within the coarse
AP and PD classes used in this analysis. in other words, it is not unreasonable to
imagine that, on average, the bin of male professional astronomers contains more
high-level scientists (in terms of career advancement and opportunities) than the
corresponding bin for female astronomers. This conjecture is supported by the
fact that no statistically significant difference between F and M is detected in the ST bin, within which no meaningful seniority gradient is expected. In the absence of higher resolution (or of a better parameter), it is not possible to disentangle the career level effect from genuine gender issues in the review process. By the same line of reasoning, the demonstrated presence of a career level effect for both genders implies that the
measured differences cannot be blindly and fully attributed to a systematic influence of gender in the reviewing process. The correlation between career level and success rate is by far the strongest
feature among all those examined in the TAWG analysis, the blurring effects discussed above notwithstanding.

\section{Comparison of gender effects}

At face value, male PIs have a factor 1.39$\pm$0.05 greater chance of getting time in the top rank classes (22.2\% vs. 6.0\%). The discrepancy becomes less marked when looking at the A+B+VM 
rates (34.2\% vs. 30.5\%), indicating that the disadvantage is generated by differences in the high-end tail of the grade distribution. This difference is illustrated in Figure \ref{fig:fig5}, which compares the distribution of pre-OPC grades of F and M PIs in the AP and PD career levels. The divergence is clearly visible (upper panel) and quantified by the difference $\Delta$ between the two cumulative functions: the peak difference indicates that M PI runs have a 6\% excess at grades better than 2.2 (lower panel). The difference is statistically very significant, as confirmed by the Kolmogorov-Smirnov test, and the low noise level characterising the $\Delta$ function. 

Another indicator, at the low-end edge of the distributions, is the fraction of triaged proposals, which is 22.1$\pm$0.6\% and
17.8$\pm$0.3\% for F and M PIs, respectively. The difference is 4.3$\pm$0.7\%, which exceeds the 6$\sigma$ level.
Similar results are obtained when considering the AP and PD career levels separately, while for the ST bin the distributions are statistically indistinguishable. A similar analysis on the post-OPC data shows the same dichotomy, although with a slightly smaller amplitude, hence signalling a mild smoothing effect operated by the panel discussions. This behaviour is observed for most of the parameters studied in the TAWG analysis.

\begin{figure}
\includegraphics[width=9cm]{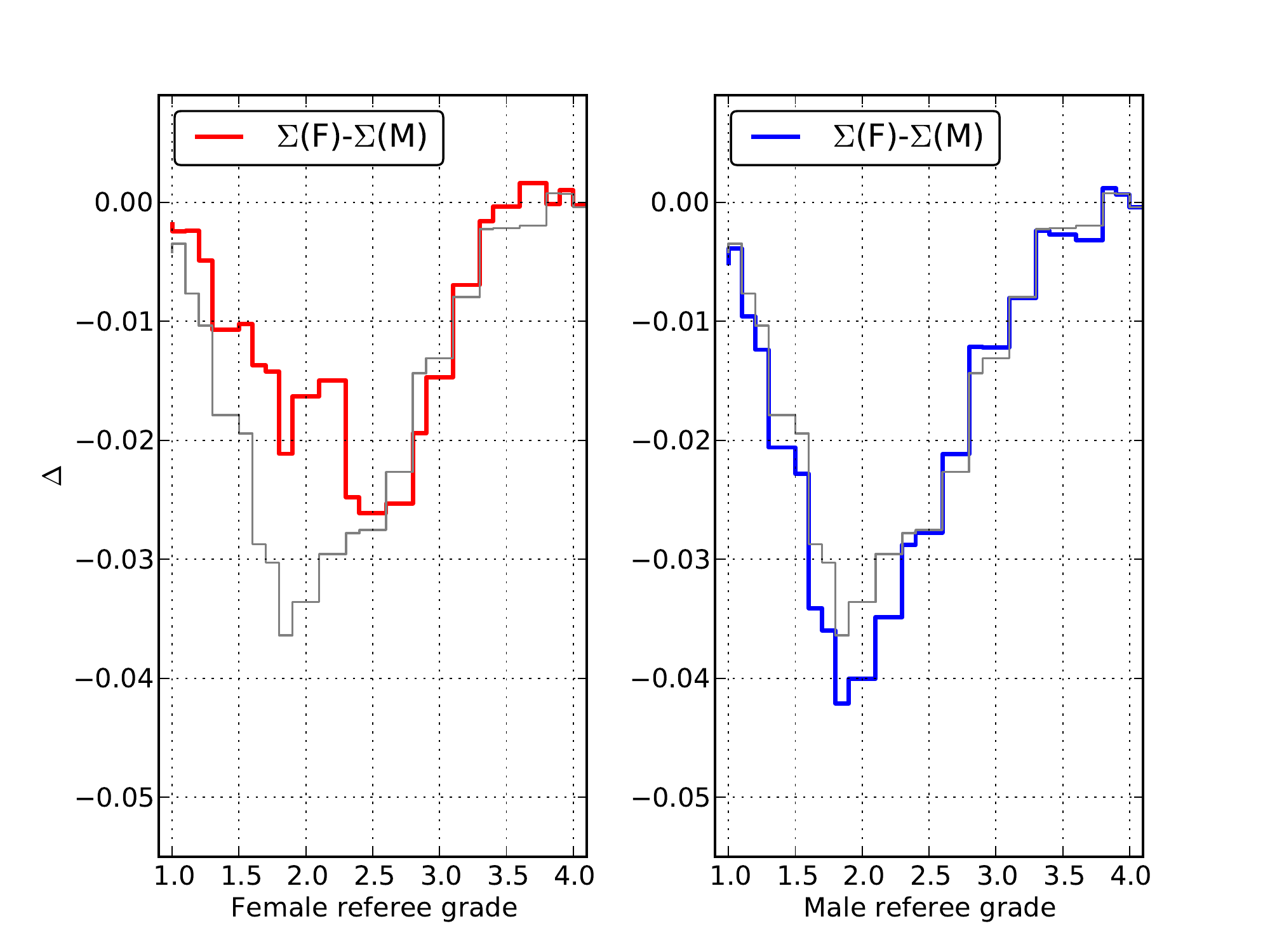}
\caption{\label{fig:fig6}Difference between the cumulative functions of grade distributions (F-M) for F (left) and M (right) referees. The grey curves trace the difference for the whole pre-OPC sample, with no referee gender distinction.}
\end{figure}

One further step in the investigation of the effect of gender systematics is the distinction between the behaviour of F vs. M referees when judging F or M PI proposals. Since the referee identity is lost in the post-OPC phase because of the way the grades are collected, this analysis is only possible for the pre-OPC grades. To compare gender-specific behaviour, the pre-OPC grades for F and M referees were extracted separately. Then, from each of the two datasets, the grade distributions for F and M PIs were derived and the $\Delta$ functions were computed. These are plotted in Figure \ref{fig:fig6}.
In both cases (F and M referees), the proposals with female PIs are disfavoured. The effect is more pronounced for M referees, with a maximum difference of 4.2\% at grade $\sim$1.8. For F referees the maximum difference is 2.6\% at grade $\sim$2.5.
When considering the whole sample, the maximum difference is 3.8\% at grade
1.8 (grey line in Figure \ref{fig:fig6}). 

The comparison between the F/M curves and the global curve gives an idea of the effect of excluding completely one of the referee genders from the panel composition. Clearly, even having an entirely F composition would not completely remove the influence of gender systematics, although it would reduce it by a factor $\approx$2. For the same reason, moving to a 50/50 balance would produce an almost unmeasurable improvement. This is in line with the results reported by \cite{reid}, who concluded that "the increased diversity on the panels has not affected the success rate of proposals with female PIs."

\begin{table}
\caption{\label{tab:tab3}Fraction of pre-OPC grades $\leq$1.9 given by referees of given gender and career level to runs requested by PIs of given gender. All values are in percent and values in parentheses are the Poisson uncertainties.}
\tabcolsep 2.4mm
\begin{tabular}{lcccc}
\multicolumn{5}{c}{\bf All referees}\\
\hline
PI & All & F & M & $\Delta$ \\
\hline
All & & 29.1 (0.3) & 26.3 (0.2) & +2.8 (0.4) \\
F & 23.5 (0.3) &  28.0 (0.6) & 23.5 (0.3) & +4.5 (0.7) \\
M & 27.8 (0.2) & 29.4(0.3) & 27.1 (0.2) & +2.3 (0.4)\\
\hline
$\Delta$ & -4.3 (0.4) & -1.4 (0.7) & -3.6 (0.4) & \\
 & & & & \\
\multicolumn{5}{c}{\bf AP referees}\\
\hline
PI & All & F & M & $\Delta$ \\
\hline
All & & 29.1 (0.3) & 26.3 (0.3) & +2.8 (0.4) \\
F & 23.8 (0.3) & 28.2 (0.6) & 23.8 (0.3) & +4.4 (0.7)\\
M & 27.9 (0.2) & 29.4 (0.4) & 27.4 (0.2) & +2.0 (0.4)\\
\hline
$\Delta$ & -4.1 (0.4) & -1.2 (0.7) & -3.6 (0.4) & \\
& & & & \\
\multicolumn{5}{c}{\bf PD referees}\\
\hline
PI & All & F & M & $\Delta$ \\
\hline
All & & 29.2 (0.8) & 23.6 (0.5) & +5.6 (0.9) \\
F  & 21.0 (1.0) & 26.9 (1.5) & 21.0 (1.0) & +5.9 (1.8)\\
M & 26.6 (0.5) & 29.9 (0.9) & 24.4 (0.6) & +5.5 (1.1) \\
\hline
$\Delta$ & -6.6 (1.1) & -3.0 (1.7) & -3.4 (1.2) & \\
\end{tabular}
\end{table}

For the sake of completeness, the same analysis was repeated on the AP and PD referee samples separately, to check whether there is any detectable gender dependency on the career level of the referee. As it turns out, there is no statistically significant difference between
the two classes. This is illustrated in Table \ref{tab:tab3}, which shows the fraction of pre-OPC grades $\leq$1.9 for the various gender/level referee combinations by PI gender. The limiting grade was chosen as approximately defining the first quartile, so that the values in the table can be considered as the fractions of top ranked runs. The $\Delta$ values are the differences between the various cumulative functions computed at the chosen limit grade; those reported in the rows correspond to the difference between PI genders, those in the last column to the difference between referee genders\footnote{An interesting aspect emerges from the $\Delta$ values reported in the last column of Table \ref{tab:tab3}: F referees tend to be slightly more lenient than M referees, meaning that their distributions tend to be shifted towards better grades. For instance, the F distribution has a nominal rejection (the fraction of grades $\geq$3.0) that is 4.0$\pm$0.4\% smaller than in the M distribution (20.6$\pm$0.3\% vs. 24.6$\pm$0.2\%), and the number of runs with grades $\leq$1.9 is 4.8$\pm$0.4\% larger than for M referees (29.1$\pm$0.3\% vs. 24.3$\pm$0.2\%).}. Although at face value PD referees show the largest deviation between F and M proposals (-6.6$\pm$1.1\%), the $\Delta$ value is only $\sim$2 standard deviations away from that derived for AP referees (-4.1$\pm$0.4\%), and therefore not much significance should be attached to it.

\section{Conclusions and actions}

The average period-by-period fraction of proposals with female PIs is $\sim$26\%, with a few percent increase during the time range covered by the present analysis. The overall fraction is $\sim$30\%. Although this value certainly underestimates the fraction in the population of potential PIs, it can be compared to the reference statistics provided by the International Astronomical Union (IAU 2016). 

\begin{table*}
\caption{\label{tab:tab4}IAU membership for ESO Member States and Chile (source IAU 2016), by number and percent. For comparison, the equivalent numbers for ESO PIs are also presented.}
\tabcolsep 3.8mm
\begin{tabular}{lccc|cc|ccc|cc}
\hline
 & \multicolumn{3}{c}{Number of IAU members} & \multicolumn{2}{c}{\% of IAU members} &
\multicolumn{3}{c} {Number of ESO PIs} & \multicolumn{2}{c}{\% of ESO PIs} \\
\hline
             & F & M & Total & F & M & F & M & N & F & M \\
Country  &    &     &          & (\%) & (\%) & & & & (\%) & (\%) \\
\hline
Austria &	13&	53&	66&	19.7&	80.3&	21&	35&	56&	37.5&	62.5\\
Belgium&	27&	119&	146&	18.5&	81.5&	20&	50&	70&	28.6&	71.4\\
Chile	 &19&	100&	119&	16.0&	84.0&	54&	169&	223&	24.2&	75.8\\
Czech Republic	 &18&	106&	124&	14.5&	85.5&	6&	18&	24&	25.0&	75.0\\
Denmark &	12&	73&	85&	14.1&	85.9&	13&	28&	41&	31.7&	68.3\\
Finland &	13&	67&	80&	16.3&	83.8&	6&	32&	38&	15.8&	84.2\\
France &	219&	619&	838&	26.1&	73.9&	106&	261&	367&	28.9&	71.1\\
Germany &	82&	594&	676&	12.1&	87.9&	167&	367&	534&	31.3&	68.7\\
Italy &	174&	494&	668&26.0&	74.0&	108&	163&	271&	39.9&	60.1\\
Netherlands &	25&	202&	227&	11.0&	89.0&	45&	96&	141&	31.9&	68.1\\
Poland &	27&	132&	159&	17.0&	83.0&	6&	18&	24&	25.0&	75.0\\
Portugal &	16&	52&	68&	23.5&	76.5&	13&	29&	42&	31.0&	69.0\\
Spain &	77&	300&	377&	20.4&	79.6&	70&	156&	226&	31.0&	69.0\\
Sweden &	20&	118&	138&	14.5&	85.5&	14&	34&	48&	29.2&	70.8\\
Switzerland &	16&	120&	136&	11.8&	88.2&	26&	61&	87&	29.9&	70.1\\
United Kingdom &	100&	604&	704&	14.2&	85.8&	130&	394&	524&	24.8&	75.2\\
\hline
All &	858 &	3753	 & 4611 &	18.6&	81.4&	805&	1911	&2716&	29.6&	70.4\\
\end{tabular}
\end{table*}

The values for the ESO Member States are presented in Table \ref{tab:tab4}. These show large fluctuations but, on average, they are significantly smaller than the above PI fraction (only the two largest values, derived for France and Italy, provide an almost exact match): the ESO Member State IAU female membership is 18.6\% which, in turn, is larger than the overall IAU fraction (16.9\%). For comparison, Table \ref{tab:tab4} also presents the equivalent numbers for ESO PIs: with very few exceptions, the F fractions are larger than the corresponding IAU values. Therefore, although certainly far from parity, the PI community is more balanced than the IAU, which is known to suffer from a strong selection bias in its membership (see for instance \cite{ces}). In his study on proposal selection at HST (covering 11 cycles between 2001 and 2013), \cite{reid} reported a female fraction growing from 19\% to 24\%, very similar to the results presented in this paper.
Despite the pro-active attitude that characterises the OPC recruiting process, the gender fraction reached in the panels at most matches that observed in the
PI distributions. This fraction decreases typically below 15\% when considering the membership of the OPC-proper, the reason being that it is more difficult to find senior female scientists willing to serve. This common problem has to do with the relatively low number of F scientists at higher career levels. Since the PI community is affected by the same imbalance, this also has a strong impact on the overall success rate by gender, as is clearly shown by the data.

Despite the coarse classification available for this study, the limitations posed by its static nature and the caveats that accompany it, the success rates shows a very marked correlation with the career level. This correlation is far stronger than any other, including that related to gender. Interestingly, this dependence is also visible in the data published by \cite{reid} for Cycles 19 and 20 (see his Figure 9), although the noise level is larger because of the smaller sample. For these reasons, any gender analysis that blindly compares F to M success rates without taking the career effect into account would mix two different issues. As a consequence, such an approach leads to a considerable overestimate of the influence of gender attributable to the proposal review process, while there certainly is an issue in the underlying population.

\begin{figure}
\includegraphics[width=9cm]{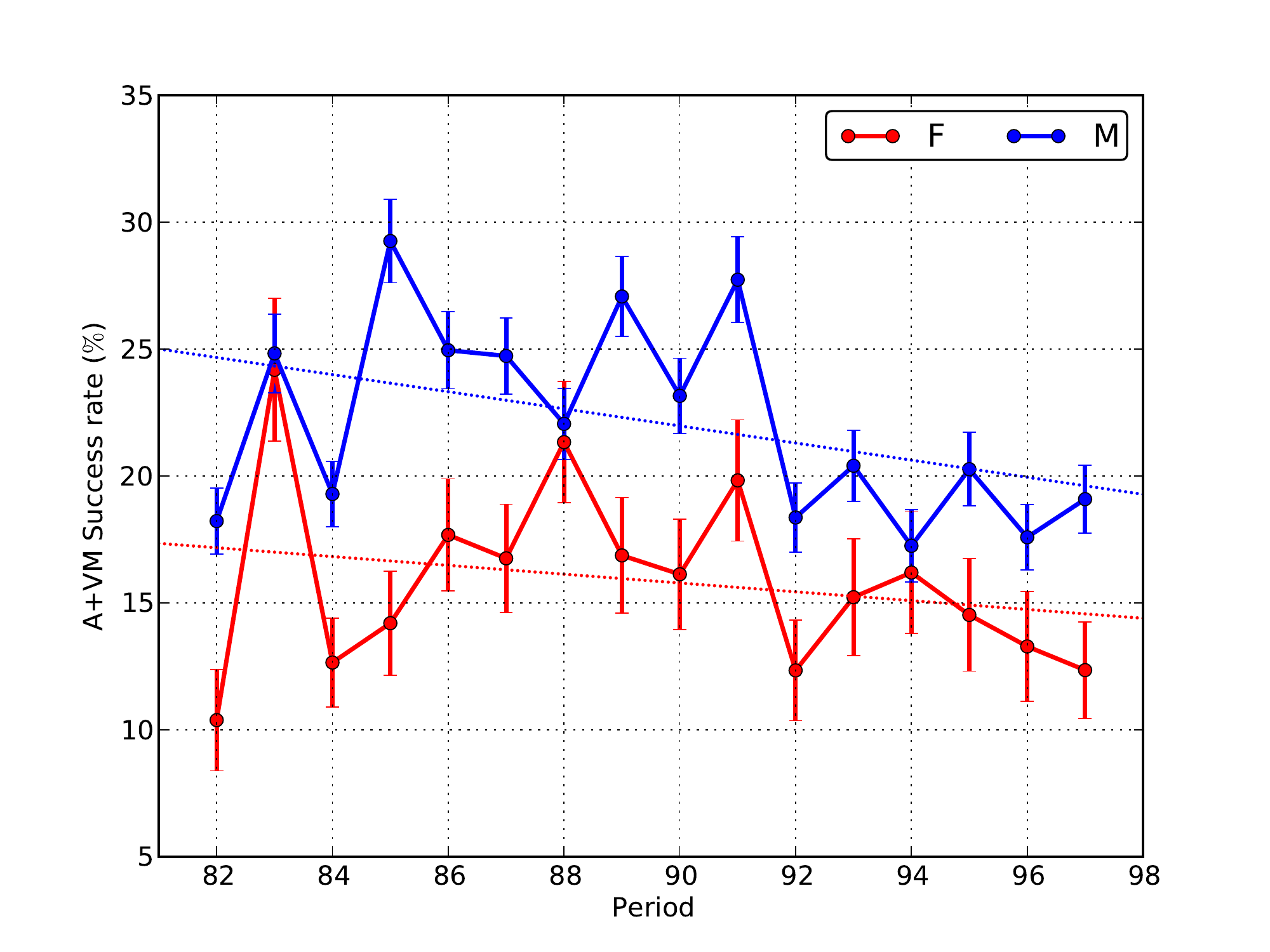}
\caption{\label{fig:fig7}Success rates for proposals at rank A and in Visitor Mode by PI gender as a function of time.}
\end{figure}

The fact remains that F success rates are consistently lower than M rates. Trending against time shows that this divergence is systematic (see Figure \ref{fig:fig7}), with a mild indication of improvement over the time span studied. The disadvantage for female PIs is similar to that derived for HST in the Cycle range 11 to 20 (\cite{reid}, Table 1): their probability of getting time
is 20-30\% lower than for their male counterparts. Although not included in the original TAWG study, an analysis of the acceptance rate of Large Programmes (LP) over the same period range (320 proposals) reveals that the discrepancy is even larger. The F fraction of submitted LPs is smaller (21.6\% vs. 25.1\% for Normal and Short programmes), and even more so is the acceptance rate (17.4\% vs. 24.7\% for M PIs), with M PIs having about 40\% greater probability of getting time through this high-impact channel.

On account of the above arguments,it is clear that this F/M difference cannot be fully attributed to the review process itself. Nevertheless, the fact that there is a significant difference between the grade distributions of F and M referees, when they rank proposals submitted
by F and M PIs, confirms that, to some extent, the reviewing process does introduce extra gender differentiation, and hence cannot be fully absolved from the charge of unequal gender treatment (be it conscious or unconscious).

ESO will continue to monitor the gender fractions in the time allocation process, will consider possible actions to mitigate the measured effects and intensify its efforts to raise awareness among the OPC panel members. However, to allow for more accurate and thorough studies, the PIs and the co-Is of proposals will
be asked to provide gender information in their UP profiles, together with a more robust career level indicator, like the year of PhD. This is fully in line with what is done at other inter-governmental facilities, and also outside of science.

While resolving the gender issue in the scientific environment requires a large and coordinated effort involving the whole community, addressing it in the more specific context of proposal evaluation is within ESO's reach. A number
of possible counter-measures can be devised, ranging from raising the awareness of reviewers to more aggressive actions, such as making the proposals anonymous (see for instance the blind audition approach used in orchestras; \cite{goldin}). While the first action is certainly necessary (and some steps have already been taken at ESO, following practices already in place at other major scientific facilities), more radical solutions need to be carefully evaluated, because they can introduce other subtle effects that would be even more difficult to quantify. This delicate topic is
discussed by \cite{reid}, who examines various possibilities, all related to the
level of information about the proposing team made available to the reviewers, and its possible implications. ESO may consider implementing the changes being tested at HST, which go along the lines of progressively obfuscating the applicants' identity (Reid 2016, private communication), possibly after the effects of such actions are statistically quantified.

\acknowledgements

I am very grateful to Elisabeth Hoppe, for painstakingly deducing the gender of more than 3000 scientists. Helpful discussions on gender issues in astronomy with Francesca Primas are also acknowledged. Special thanks go to the members of the Time Allocation Working Group, for all their support and collaboration. In particular, I am grateful to Neill Reid for the very fruitful exchange of ideas on telescope time allocation matters. Finally, I wish to thank Rob Ivison and Tim de Zeeuw for suggesting the use of a seniority indicator and for kindly reviewing the manuscript, which greatly improved its quality.

\end{document}